\documentclass[11pt]{article}
\usepackage{moriond,epsfig}
\usepackage{wrapfig}
\usepackage{amsmath}
\usepackage{amssymb}
\usepackage{hyperref}

\newcommand{\journal}[4]{{\em #1} $\;${\bf #2} $\;$ (#3) $\;$ #4}

\newcommand{\prl}{\journal {Phys. Rev. Lett.}}

\newcommand{\app}{\journal {Acta Phys. Pol.}}

\newcommand{\apj}{\journal {Astrophys. J.}}

\newcommand{\PRD}{\journal {Phys. Rev. D}}

\def\be{\begin{equation}}
\def\ee{\end{equation}}
\def\bc{\begin{center}}
\def\ec{\end{center}}
\def\bea{\begin{eqnarray}}
\def\eea{\end{eqnarray}}

\newcommand{\gev}{\ensuremath{{\mathrm{GeV}}}}

\begin{document}
\vspace*{4cm}
\title{The IceCube Cosmological 
Connection: \\Status and prospects 
of the polar neutrino observatory}

\author{Mathieu Ribordy\\
for the IceCube collaboration}

\address{High Energy Physics Laboratory, EPFL, CH - 1015 Lausanne}

\maketitle\abstracts{
We report on the current construction status of the IceCube high energy neutrino observatory and possible future construction plans. With the completion of the fourth construction season in Feb. 2008, the observatory is now instrumenting half a cubic kilometer of ice, greatly increasing the horizon for high energy neutrino detection. We briefly describe physics topics related to cosmology, such as indirect searches for supersymmetric cold dark matter, for slow and relativistic magnetic monopoles, GZK neutrinos and violation of Lorentz invariance or Equivalence Principle. 
It is anticipated that upon completion the new detector will vastly increase the sensitivity and extend the reach of AMANDA to higher energies. 
}

While IceCube is primarily designed for the resolution of the obscure origin of the charged cosmic rays by observing the neutrino sky, it also offers potential for various particle physics and cosmology related topics. We specifically discuss a quest for ultra high energy neutrinos and searches for deviations of the atmospheric neutrino flux from violations of the Lorentz invariance, magnetic monopoles and cold dark matter.

\section{Performance of the IceCube detector}\label{sec:det}
After four successful construction seasons, the first half of the IceCube detector is deployed. It will eventually consist of an array of 80 strings buried at depths between 1.5 and 2.5 km under the South Pole ice cap and instrumenting a 1 km$^3$ hexagonal volume encompassing the existing AMANDA-II detector. The end of the deployment is planned for 2011.
The 1 km long strings are equipped with 60 digital optical modules each, arranged in a triangular lattice with 125 m edge. These modules are autonomous sensitive devices consisting of a photomultiplier and electronic boards ensuring signal digitization and communication with the ground surface. Prospective analyses~\cite{henrike} have shown that sensitivities will improve by more than an order of magnitude over the existing AMANDA-II detector. However, due to a sparser instrumentation, a slight shift of the energy threshold towards higher energies is expected.
During the course of construction, an ever increasing deployment pace has been achieved and the event rate has been continuously increasing. 
Ultimately, $\approx 300$ atmospheric neutrino-induced muon events will be recorded per day.
The deployment of 6 additional strings in the lower part of the in-ice array could start next season. This deep core (including some of the original IceCube strings) with reduced string spacing would extend the reach of IceCube towards lower energy. The opportunity for a 100 km$^3$ ultra high energy radio-acoustic hybrid extension is also investigated. Both extensions are of specific interest for the topics presented below.

\section{GZK neutrinos}\label{sec:gzk}
\begin{wrapfigure}{r}{5.5cm}
\mbox{\includegraphics[angle=0,width=0.33\textwidth]{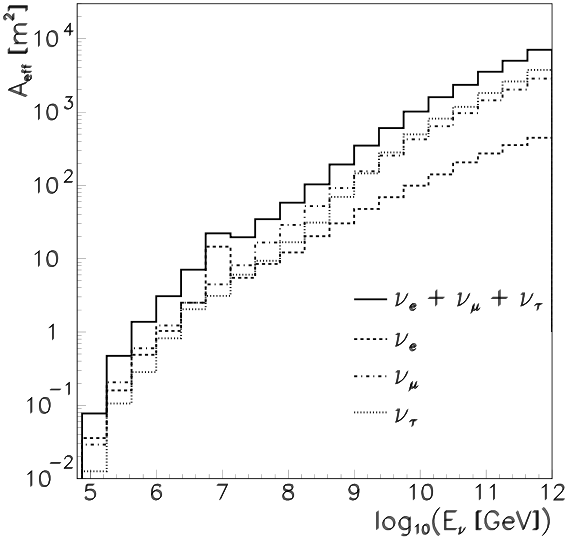}}
\caption{$\langle A_\mathrm{eff} \rangle_\theta$.\hfill~}
\label{fig:uhe}
\end{wrapfigure}
Ultra high energy cosmic rays undergo energy damping above $\approx 10^{20}$ eV/nucleon over cosmologically short distances as they interact with the cosmic microwave background, photoproducing pions.
The resulting GZK neutrino flux, however, remains unaffected and constitutes therefore a precious information carrier. Characterizing this flux would allow the recovery of most of the lost information and eventually reconstitute source injection spectra along with source cosmological evolution.
Analyzing data taken with AMANDA between 2000 and 2002\cite{lisa} in search for these ultra high energy neutrinos nevertheless remains a challenge due to systematic uncertainties at various levels (ice properties, absolute detector sensitivity, primary cosmic ray flux normalization and composition, interaction cross sections) which must be accounted for. The main background to fight consists of high energy muon bundles. A 90\% C.L. upper limit $E^2 \Phi = 2.4 \cdot 10^{-7}$ GeV cm$^{-2}$ s$^{-1}$ sr$^{-1}$ for a $E^{-2}$ neutrino signal in the range $2 \cdot 10^5$ GeV $<E<$ $10^9$ GeV has been derived after the analysis revealed no excess ($2$ events in the final sample for an expected background $<2.1$). The angle-averaged effective area $\langle A_\mathrm{eff} \rangle_\theta$, shown Fig.~\ref{fig:uhe} corresponds to an expectation of 0.06 GZK neutrino events\cite{stanev}. The IceCube detector, with its yearly growth of the instrumented volume may however be close catching first GZK neutrinos for average cosmogenic flux estimates by the end of 2010.
The intertwine of astrophysics, particle physics and cosmology renders this quest a particularly interesting challenge and new information would help constrain possible scenarios. Means for surmounting the limited sensitivity of IceCube in this UHE range are being considered with ongoing radio-acoustic detection techniques R\&D activities.

\section{Equivalence principle and Lorentz invariance}\label{sec:vli}
\begin{wrapfigure}{r}{6cm}
\mbox{\includegraphics*[scale=.44]{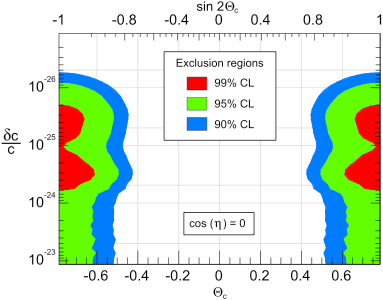}}
\caption{$\frac{\delta c}{c}$ exclusion region w.r.t. the $\Theta_C$ MAV mixing angle.}
\label{fig:vli}
\end{wrapfigure}
Many attempts to build a quantum gravity theory require the deformation or the breaking of the Lorentz symmetry and predict non trivial modifications of space-time symmetries at the Planck scale.
This may lead to modifications of the dispersion relations and consequently the violation of the Lorentz invariance. 
Modified kinematic thresholds in scattering and decay processes and species-dependent maximum attainable velocities (MAV) are among possible associated consequences. The latter may induce neutrino oscillation differing in pattern from the conventional mass-induced oscillations.
The modification of the dispersion relation, retaining the validity of the energy-momentum conservation law, is a simple kinematic framework to introduce a violation: $E^2 = m_i^2 + (1+f_i)p^{2}$, where $i$ denote the energy eigenstates and where the $f_i \ll 1$ depend on the species. 
The oscillation probability is therefore $P_{\nu_\mu \rightarrow \nu_\mu} = 1-\sin^2{2\theta_v}\sin^2{(\delta v EL/2)},$
where $\theta_v$ is the MAV mixing angle to the flavor eigenstates and $\delta v$ the fractional MAV difference. Note that oscillations also occur if the neutrino mass vanishes and the $EL$ dependence of the argument instead of the usual $L/E$ in standard oscillations. Thus, the exotic term competes with the classical term and may dominate at high energy,
\begin{wrapfigure}{r}{5.5cm}
\includegraphics[angle=0,width=0.34\textwidth]{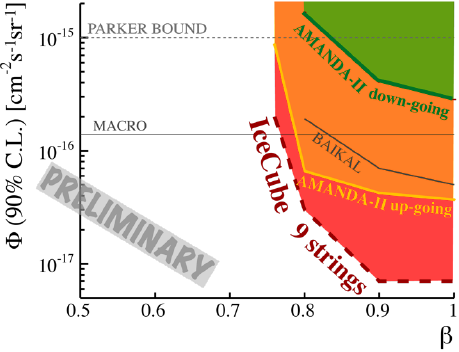}
\caption{Relativistic MM flux limit and sensitivity w.r.t. $\beta$}
\label{fig:mm1}
\end{wrapfigure}

\noindent leading to distortions of the angular and energy distributions relative to pure mass-induced oscillations. Distribution discrepancies beyond statistical and systematic errors would suggest a violation of the Lorentz invariance. Alternatively, this would help constrain the allowed parameter space ($\delta v$, $\theta_v$ and a phase $\eta$).
According to~\cite{vli-halzen}, IceCube will collect a unique sample of about 240k atmospheric neutrinos with energies above 50 GeV by the end of 2011. This will permit for testing the Lorentz invariance down to $\delta v \lesssim 10^{-27}$ in the particular case of coinciding mass and MAV eigenstates. 
Here we report on a constraint resulting from the analysis of four years of data taking with the AMANDA detector (2000-2003)\cite{vli}: For various parameter (phases, $\delta m^2$, mixing angles) values, the exclusion regions w.r.t. the mixing angle $\theta_v$ are set (Fig.~\ref{fig:vli}). This analysis constrains a violation of the Equivalence Principle alike, swapping $\delta v_i$ with $\delta \Phi_i$, the coupling to the gravitational potential of species $i$.


\section{Magnetic monopoles}\label{sec:mm}
\begin{wrapfigure}{r}{5.5cm}
\includegraphics[angle=0,width=0.35\textwidth]{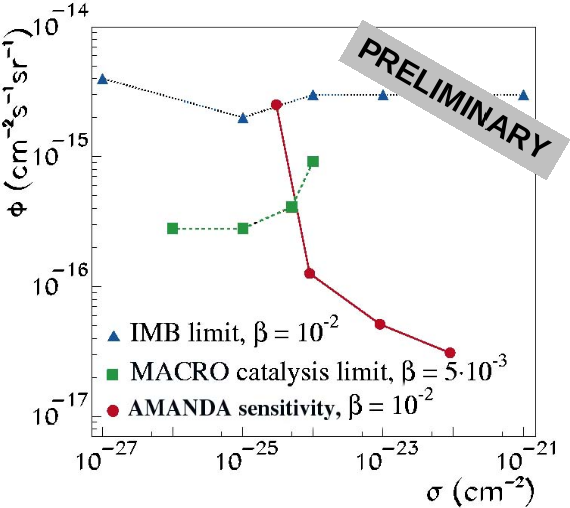}
\caption{Slow MM flux limits (IMB, MACRO) and expected sensitivity w.r.t. the catalysis cross section, $\beta=10^{-2}$.}
\label{fig:mm2}
\end{wrapfigure}
Magnetic monopoles (MM) are topological defects produced during phase transition in the early universe. They may be stable and have survived until now. With AMANDA-II and the 9 string IceCube configuration (IC-9), searches for relativistic and slow ($\beta=10^{-2}$) MM were performed.

The relativistic MM search methodology applied to the sample depends on the MM mass: At masses in excess of $m_\mathrm{MM} \gtrsim 10^{11} \,\gev$, MM can travel upward through the detector, providing a unique signature of a bright event which cannot be mimicked by a possible UHE neutrino event, the latter being screened by Earth. Limits on a MM flux set with the AMANDA detector\cite{mmm} are shown Fig.~\ref{fig:mm1} as well as a promising IC-9 projected sensitivity\cite{mmm2} for future analyses with the further deployed IceCube detector.
At lower masses, the conventional background of very bright events has to be dealt with, the limit which can be set on the MM flux with AMANDA is in this case not as constraining.
Slow monopoles on the other hand are exhibiting the specific signature of a long duration event, far exceeding the data acquisition system capabilities to record its complete trajectory.
A slow monopole crossing the detector is catalyzing proton decays along its trajectory and 
Cherenkov light is emitted from the frequent $E \approx m_\mathrm{p}$ EM showers. A preliminary sensitivity on the slow monopole flux was set w.r.t. the catalysis cross section\cite{smmm}, suggested to have strength of around strong interaction cross section. Fig.~\ref{fig:mm2} shows the complementarity of this search for a range of cross sections not covered by the MACRO experiment.


\section{Supersymmetric cold dark matter}\label{sec:cdm}
One of the most urgent cosmological problem comes from the non observation of $\approx$80\% of the matter content of the universe. High energy neutrino telescopes have the potential to explore this matter further. In the case of non baryonic cold dark matter in the form of the lightest neutralino in the framework of the minimal supersymmetric extension of the standard model (MSSM): neutralinos in the halo scattering off ordinary matter may eventually be gravitationally trapped in macroscopic objects. Subsequent scatterings then reduce their velocity resulting in

\begin{wrapfigure}{r}{5.8cm}
\mbox{\includegraphics[angle=0,width=0.36\textwidth]{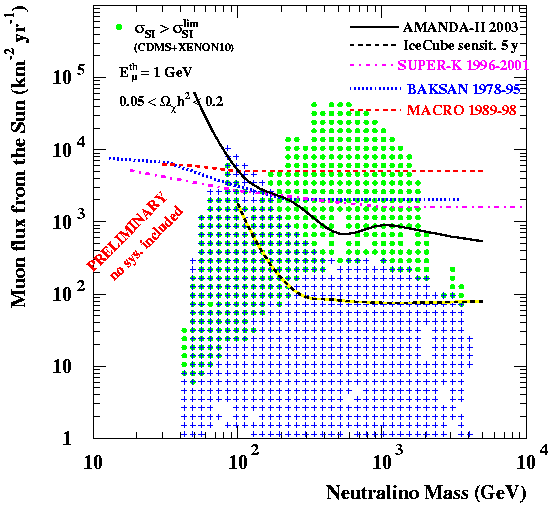}}
\caption{Preliminary AMANDA-II flux limit and 5 yr IceCube sensitivity $\Phi_\mu$ vs $m_\chi$ for various MSSM models.\hfill~}
\label{fig:wimp}
\end{wrapfigure}
\noindent  their accumulation in  the  core of  these  objects over  astronomical  times,
where they annihilate pairwise producing neutrinos that can be detected.

Preliminary results of a search performed with the AMANDA-II detector, analyzing the data taken in 2003, in terms of muon flux $\Phi_\mu$ w.r.t. the neutralino mass $m_\chi$ (note that this may worsen a bit when systematic uncertainties are included) is shown Fig. \ref{fig:wimp}: Dots (crosses) are for various MSSM models which are (not yet) excluded by direct search experiments\cite{xenon,cdms}.
Once 5 years of data will be collected with the completed IceCube, the WIMP sensitivity is expected to explore the MSSM parameter space beyond the reach of the current direct CDM search experiment. Also note that nuclear recoil and indirect CDM search experiments are not equivalent. For an unevenly distributed CDM halo throughout the galaxy or a neutralino mixture with large relative spin-dependent cross section, the latter have a detection potential which remains intact.


\section*{Conclusions and perspectives}
In conclusion, IceCube's unprecedented sensitivity may provide access to exotic physics (neutralino cold dark matter, monopoles, violation of the founding principle of (general) relativity). From the astrophysical perspective, while IceCube may soon reveal the first high energy extraterrestrial neutrinos within a few years, contributing to unveil the cosmic ray origin mystery, R\&D efforts toward a hybrid extension of IceCube are to be pursued: The GZK neutrino flux characterization would give insights into cosmological source evolution and source spectra from the partial recovery of the degraded information carried by the ultra high energy cosmic rays.

\section*{Acknowledgments}
The author is supported by the Swiss National Research Foundation.

\end{document}